\documentclass[10pt,prl,aps,superscriptaddress,twocolumn,longbibliography,showpacs]{revtex4-1}
\usepackage{amssymb,amsmath}
\usepackage{graphicx}
\usepackage{natbib}
\usepackage{hyperref}
\usepackage[utf8]{inputenc}

\begin{document}

\newcommand{\e}{{\rm e}}
\newcommand{\rmi}{{\rm i}}
\renewcommand{\Im}{\mathop\mathrm{Im}\nolimits}
\newcommand{\red}[1]{\textcolor{red}{#1}}
\newcommand{\blue}[1]{{\color{blue}#1}}

\newcommand{\refPR}[1]{[\onlinecite{#1}]}
\newcommand{\cra}[1]{\hat{a}^{\dag}_{#1}}  
\newcommand{\ana}[1]{\hat{a}_{#1}^{\vphantom{\dag}}}         
\newcommand{\num}[1]{\hat{n}_{#1}}         
\newcommand{\bra}[1]{\left|#1\right>}      
\newcommand{\eps}{\varepsilon}      
\newcommand{\om}{\omega}      
\newcommand{\kap}{\varkappa}      

\newcommand{\spm}[1]{#1^{(\pm)}}
\newcommand{\skvk}[2]{\left<#1\left|\frac{\partial #2}{\partial k}\right.\right>} 
\newcommand{\skvv}[2]{\left<#1\left|#2\right.\right>}
\newcommand{\df}[1]{\frac{\partial #1}{\partial k}}
\newcommand{\ds}[1]{\partial #1/\partial k}

\title{Topological edge states of bound photon pairs}
\author{Maxim~A.~Gorlach}
\affiliation{ITMO University, Saint Petersburg 197101, Russia}
\author{Alexander N. Poddubny}
\affiliation{ITMO University, Saint Petersburg 197101, Russia}
\affiliation{Ioffe Institute, Saint Petersburg 194021, Russia}
\email{poddubny@coherent.ioffe.ru}

\begin{abstract}
We predict the existence of interaction-driven edge states of  bound two-photon quasiparticles in a dimer periodic array  of nonlinear optical cavities. 
Energy  spectrum of photon pairs is dramatically richer than in the noninteracting  case or in a simple lattice, featuring  collapse and revival of multiple edge and bulk modes as well as edge states in continuum. Despite the unexpected  breakdown of the Zak phase technique and the edge mixing of internal and center-of-mass motion we link the edge state existence to  the two-photon quantum walk graph connectivity, thus uncovering the  topological nature of the many-body problem in complex lattices.
\end{abstract}

\maketitle

Nonlinear and many-body phenomena in condensed matter physics and optics are currently in the focus of research interest due to the wide range of opportunities including realization of strongly correlated photon gases, implementation of polariton superfluidity and formation of solitons and vortices~\refPR{Carusotto}. One of such striking interaction-induced effects  has recently been experimentally observed by Winkler et~al.~\refPR{Winkler}. In the presence of a repulsive interaction two bosons can form a bound pair propagating as a single quasiparticle in an optical lattice \cite{Winkler,Valiente}. Such repulsively bound pairs further termed as doublons do not have direct analogues in traditional condensed matter systems being of fundamental interest for the many-body physics  and quantum information.

Search for the doublon edge states~\refPR{Flach,Pinto,Longhi} has revealed their absence in a simple periodic lattice~\refPR{Flach}. Tamm edge states of doublons may arise similarly to the single-particle case if the  lattice has an edge defect~\refPR{Longhi}. However, the existence of topological doublon edge states as well as the calculation of topological invariants for composite particles remain open problems so far.

\begin{figure}[b]
    \begin{center}
    \includegraphics[width=0.95\linewidth]{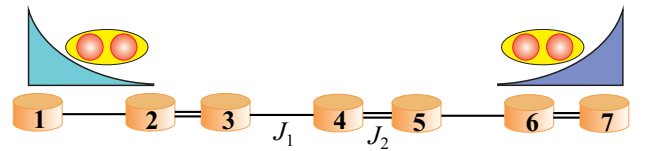}
    \caption{Dimer lattice  of nonlinear cavities with the tunneling constants $J_1$ and $J_2$. Two-photon edge states are sketched.}
    \label{fig:System}
    \end{center}
    \end{figure}

Inspired by the recent advances in topological  photonics~\refPR{Lu2014,Khanikaev,Ma,Slob} and quantum optics \cite{Greentree,Hartmann2006}, we investigate the edge states of doublons in a {\it dimer lattice} of identical optical cavities with  a Kerr-type nonlinearity and two alternating tunneling constants, $J_1\ne J_2$, see Fig.~\ref{fig:System}.  This is a many-body generalization of the 
Su-Schrieffer-Heeger  (SSH) model, considered a simplest  example for  the topological edge states of photons \cite{Schomerus:13} and plasmons \cite{Ling2015,Zhu2015}. Topological transitions in somewhat similar {\it classical} nonlinear systems have been recently predicted in Refs.~\cite{Hadad,Solnyshkov}.  
Ground state of  the many-body SSH model has been studied in Ref.~\cite{Grusdt2013}.
To the best of our knowledge, the {\it two-particle excitations} of the interacting SSH model are fully uncharted.  Here, we present a rigorous numerical diagonalization for the finite system, accompanied by an exact analytical solution for the bulk two-photon states and by a condition for the edge states, based on the topology of the  two-photon quantum walks in the bulk. While the  quantum walks have been extensively studied \cite{Aharonov2001,Omar2006,Lahini2012,Manouchehri2013physical} and even successfully used to describe the noninteracting topological model \cite{Demler2010}, we are not aware of their direct applications to characterize the edge states in interacting systems.

We employ the Bose-Hubbard-type Hamiltonian~\cite{Essler,Grusdt2013} 
\begin{equation}\label{eq:Hamiltonian}
\begin{split}
& \hat{H}=\omega_0\,\sum\limits_{m}\num{m}+U\,\sum\limits_{m}\,\num{m}\,(\num{m}-1)\\-
& J_1\,\sum\limits_{m}\,\left(\cra{2m-1}\,\ana{2m}+\cra{2m}\,\ana{2m-1}\right)\\-
& J_2\,\sum\limits_{m}\,\left(\cra{2m}\,\ana{2m+1}+\cra{2m+1}\,\ana{2m}\right)\:,
\end{split}
\end{equation}
where $\cra{m}$ and $\ana{m}$ are the photon creation and annihilation operators for   the $m$-th cavity, $\num{m}=\cra{m}\,\ana{m}$ is the photon number operator, $J_{1,2}$ are the tunneling constants, $U$ is the interaction strength  and $\omega_0$ is the cavity eigenfrequency ($\hbar=1$). We assume fixed photon polarization. 

\begin{figure}[t]
    \begin{center}
    \includegraphics[width=0.95\linewidth]{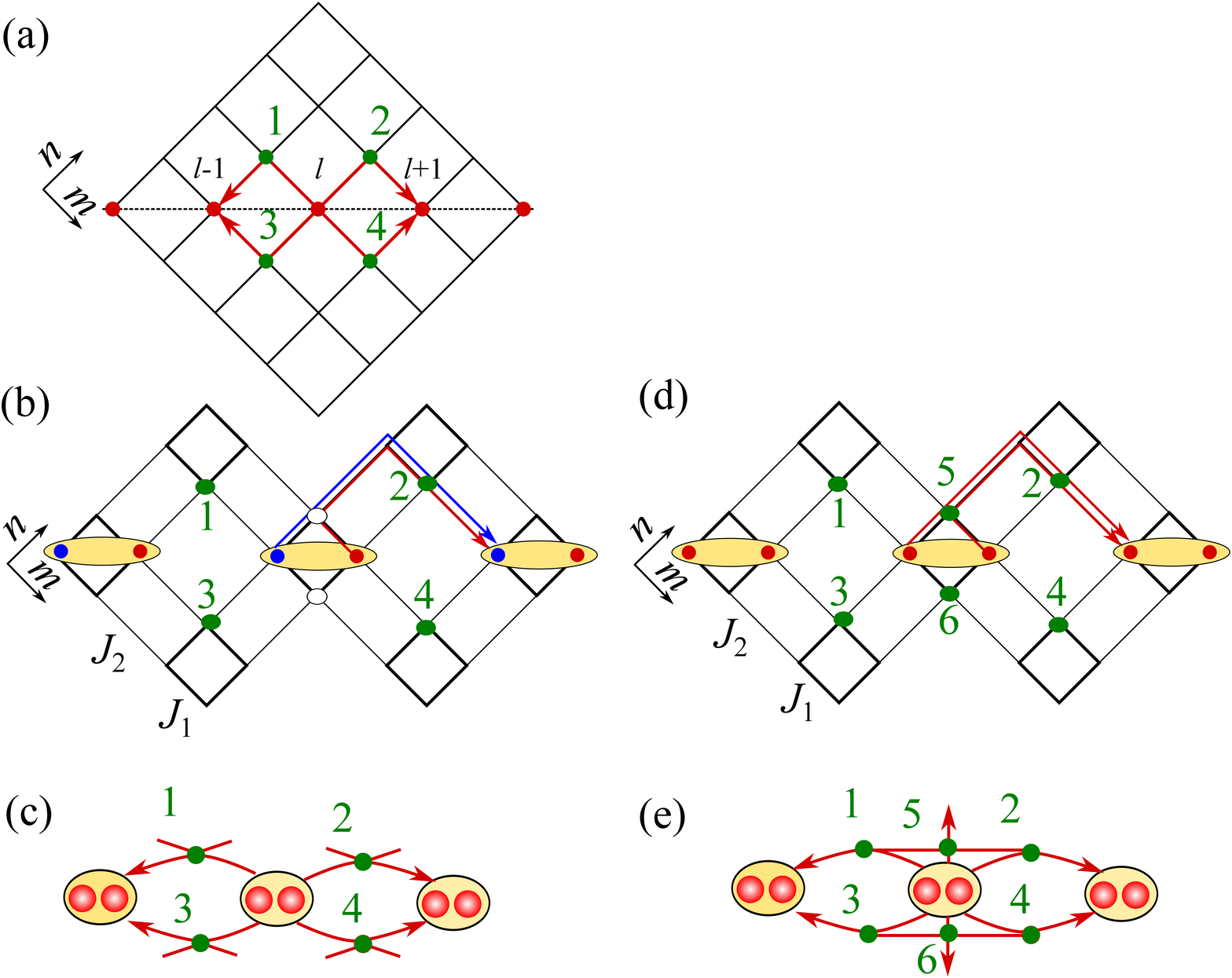}
    \end{center}
    \caption{ Graphic representation of the photon pair Hamiltonian (a,b,d) and quantum walks graphs (c,e)
    in the trivial (left column) and nontrivial (right column) cases. Panel (a) corresponds to $J_1=J_2$, panels (b) and (d) to the odd and even modes for $J_1>J_2$, respectively.
 Sites with $n=m$ have in     (a,b,d) the energy  $2U$, other sites have zero energy. 
    }
    \label{fig:2D}
    \end{figure}

Since the Hamiltonian Eq.~\eqref{eq:Hamiltonian} commutes with the operator  $\sum_{m}\,\num{m}$,  the total number of photons is conserved and  the two-photon state can be searched as
$\bra{\psi}=\sum_{m,n}\,\beta_{mn}\cra{m}\cra{n}|0\rangle\:,$  with $\beta_{mn}=\beta_{nm}$.
Substituting the wave function  into the Schr\"odinger equation
$
\hat{H}\bra{\psi}=(\eps+2\omega_0)\,\bra{\psi}
$
we obtain a linear system of equations to determine the  coefficients $\beta_{mn}$ and the energy $\eps$. Thus, the  one-dimensional two-photon problem is equivalent to  a  two-dimensional (2D) single-particle problem~\cite{Longhi}. The corresponding 2D lattice is illustrated in Fig.~\ref{fig:2D}(a) for $J_1=J_2$ and in Figs.~\ref{fig:2D}(b,d) for $J_1<J_2$. The $m$ and $n$ coordinates are just the coordinates of first and second photon. The  links represent the tunneling amplitudes $J_{1,2}$; the energies for the diagonal sites with $m=n$ are equal to $2U$ (two photons in the $n$-th cavity), the off-diagonal sites have zero energies (space-separated photon pair).
Next, we show how the existence of the doublon edge states follows from  the connectivity of the two-photon tunneling pathways in this 2D lattice and verify the result by the full diagonalization.

We first reexamine the case $J_1=J_2\equiv J$, where the doublon edge states are absent~\cite{Flach,Pinto}, in the  limit of strong interaction $U\gg J$. In the zeroth order approximation in $J/U$ all the doublon states have the energy $\eps\approx 2U$, they are degenerate and the two photons are pinned to the same cavity [dotted diagonal in Fig.~\ref{fig:2D}(a)].  The tunneling couples doublons at different cavities as well as shifts their energies. The resulting effective Hamiltonian for doublons has the general form 
\begin{equation}
\hat{H}_{\rm eff}=\sum\limits_{l=1}^N(\eps_0+\delta_l)\hat d_l^\dag \hat d_l^{\vphantom{\dag}}+t\sum\limits_{l=1}^{N-1}
(\hat d_l^\dag \hat d_{l+1}^{\vphantom{\dag}}+{\rm H.c.}),\label{eq:Heff}
\end{equation}
where $\eps_0=2U$ and $\hat d_l=\frac{1}{2}\hat a_l^2$. The effective two-photon tunneling amplitude  $t$ and  the two-photon energy blueshift $\delta_l$ can be expanded in the perturbation series in $J/U$. We resort to the nearest-neighbor coupling for doublons. In this case the tunneling constant can be calculated by means of the second order  perturbation theory,   $t=J^2/U$~\cite{Flach}. This corresponds to the two photons tunneling to the adjacent cavity directly one after another. The corresponding tunneling pathways are shown in the two-photon quantum walk graph of Fig.~\ref{fig:2D}(a) by the red arrows.  The loops linking the site $l$ to itself, i.e. the path $l\to 2\to l$ in  Fig.~\ref{fig:2D}(a), contribute to the  energy shifts $\delta_l$.

The existence of edge states in the model Eq.~\eqref{eq:Heff} depends
on the relation between  $\delta_l$ and $t$ that, in turn, follows from  the topology of the quantum walk graph.
Two important identities hold   for $J_1=J_2$:
$
\delta\equiv 2\delta_1=\delta_2=\ldots \delta_{N-1}=2\delta_N$  and 
$\delta=2t$.  The first identity stems from the translational symmetry and the fact that the first and last cavities have twice less neighbors. The $\delta=2t$ condition means that the self-induced nonlinear blueshift of the site in the bulk  can not exceed the energy shift induced by the left and right neighbors. This condition arises because of the mirror symmetry. Namely, the mirror reflection with respect to the vertical lines 1--3 or 2--4 in Fig.~\ref{fig:2D}(a) maps each two-step path coupling nearest neighbors  to the path coupling the site to itself, e.g. $l\to 2\to l+1$ to $l\to 2\to l$. 
There exist two symmetry lines, so $\delta=2t$. In terms of the  two-photon quantum walks topology the argument above means that  the  {\it local vertex connectivity } $\kappa$, corresponding to the eigenmode, is equal to $4$. The vertex connectivity is by definition the minimal number of vertices that have to be cut from the graph to make it disconnected~\cite{West2001}. The relevant two-photon tunneling pathways in Fig.~\ref{fig:2D}(a) are schematically shown in Fig.~\ref{fig:2D}(c) with  the points 1--4 forming the minimal set of vertices to be cut. Given the conditions above, the Hamiltonian Eq.~\eqref{eq:Heff} describes a simple periodic lattice,
where the first and the last sites are detuned by the energy $\delta_2-\delta_1\equiv\delta/2$ from the middle ones. The detuning is exactly equal to the tunneling amplitude, $\delta/2=t$. Hence, the condition $\delta/2>t$ for the edge states in the 1D tight-binding model with the edge defect  [Sup.~Mat.~V] is not satisfied.  The edge states do not exist for $J_1=J_2$, although they can be enabled by an arbitrary weak perturbation~\cite{Pinto,Longhi}.

The above analysis of the  two-photon quantum walks  can be generalized to the much less trivial dimer lattice with $J_1\ne J_2$. We now assume that 
$U\gg J_1\gg J_2>0$, the number of cavities is even, $N=2M$, and the lattice ends with a strong tunneling link at both edges.   In the perturbation scheme we first take into account the strong links $J_1$ and then the weak links $J_2$. Due to the dimerization the correct doublon operators in the zeroth order in $J_1/U$ are the odd and even combinations  $\hat d^{(\pm)}_{l}=(\hat a_{2l-1}^2\pm\hat a_{2l}^2)/(2\sqrt{2})$. The effective doublon Hamiltonian still has the structure 
Eq.~\eqref{eq:Heff} with $\eps_0^{(\pm)}=2U+J_1^2/2U\pm J_1^2/2U $ and 
$N$ being replaced by  $M=N/2$. The energies $\delta_l$ and the tunneling amplitude $t$ are modified.
Since now  each doublon mode $\hat d^{(\pm)}_l$ occupies two cavities instead of one, the consistent derivation of  $\delta_l$ and $t$ requires fourth-order perturbation theory [Supplemental Materials, Sec.~IV], i.e. including all two-step and four-step pathways in Figs.~\ref{fig:2D}(b,d). 
We find that  contrary to the $J_1=J_2$ case, the condition $\delta=2t$ no longer holds in general because the local vertex connectivity $\kappa$ of the eigenmode can exceed 4.  
Indeed, for the even mode $\kappa=6$ as illustrated in Fig.~\ref{fig:2D}(d).  
The corresponding quantum walk graph [Fig.~\ref{fig:2D}(e)]  is inherently  irreducible to the graph for $J_1=J_2$ in Fig.~\ref{fig:2D}(c).
Consequently, more pathways contribute to the blueshift than to the tunneling, and one has $\delta_+-2|t_+|=(J_1J_2)^2/U^3>0$ {[Sup.~Mat.~IV]}. Since the detuning of  the bulk sites from the edge $\delta_{+}/2$  exceeds $|t_+|$, the edge state does appear.  As such, the condition $\kappa>4$ can be used to predict the presence of the edge state.

The odd mode $\hat d^{(-)}_{l}$ still has $\kappa=4$ and no edge states for strong link termination. The contributions from all the pathways crossing the points $5$ and $6$ cancel each other  due to the odd mirror symmetry, so these points are to be excluded from the graph, see Fig.~\ref{fig:2D}(b). More detailed analysis, presented in the Supplementary Materials, confirms that  the doublon edge state is formed  for the even band but not for the odd one. Its localization length  is given by $(U/J_1)^2$ and the energy is close to the lower edge of the even band $\eps_0^{(+)}$.  The crude argument for the emergence of the edge state only for the even band of doublons is that its blueshift due to the interaction is stronger in general than that for the odd mode. Hence, the energy detuning between the edge and bulk sites is larger and this facilitates the edge state formation. We stress that the two-photon state is formed at the edge with strong tunneling link, where the single-photon edge states are absent.

   \begin{figure}[t]
    \begin{center}
    \includegraphics[width=0.45\textwidth]{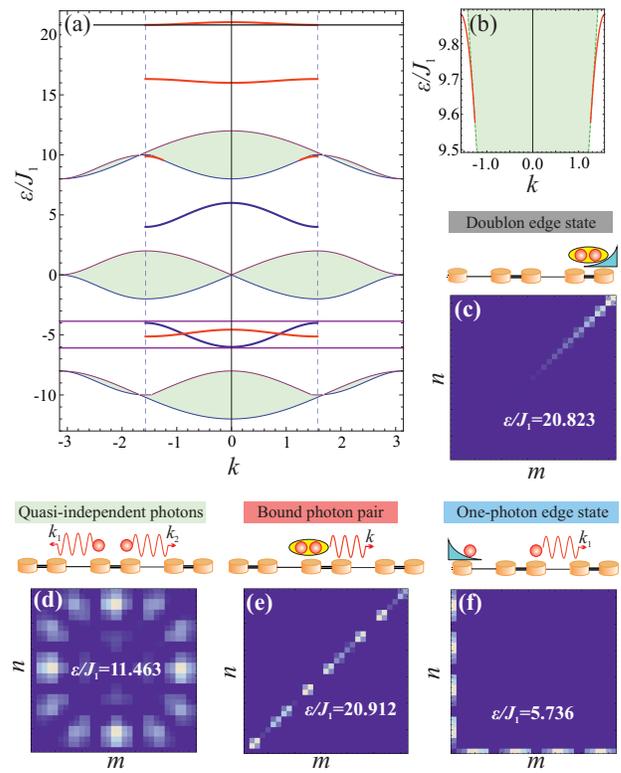}
        \end{center}
    \caption{(a) Dispersion of various types of two-photon excitations for fixed parameter values: $J_2/J_1=5$, $U=8J_1$. (b) Collapse of the doublon band. (c-f) Typical probability distributions for different types of two-photon excitations in a finite array  of $N=31$ cavities: (c) doublon edge state localized at the edge with strong or weak tunneling link [black or purple lines in panel (a)] ; (d) two quasi-independent photons (green bands); (e) bound photon pair [doublon, red line]; (f) single-photon edge state (blue line).}
    \label{fig:Distr}
    \end{figure}

The analysis above has been performed in the limit of strong interaction and focused on the doublon edge states only. Now we will discuss the whole energy spectrum for an arbitrary interaction strength.
In the infinite lattice two types of excitations are possible: (i) pairs of quasi-independent photons that move along the lattice almost without  interaction and have the energies $\eps=E_1(k_1)+E_1(k_2)$, where 
$E_1(k)=\pm\sqrt{J_1^2+J_2^2+2J_1J_2\cos 2k}$ is the single-photon energy, $k_{1,2}$ being the real Bloch wave numbers varying from $-\pi/2$ to $\pi/2$ and (ii) doublons with complex $k_1$ and $k_2$ of the form $k_1=(k-\varkappa)/2$, $k_2=(k+\varkappa)/2$, where $k=k_1+k_2$ is a real number describing the motion of photon pair as a whole, while a complex number $\varkappa$ describes the relative motion of photons confined to each other. 
 As shown in the Supplementary Materials, the {\it bulk} doublon  states can be sought in the form of modified Bethe anzatz
that 
yields exact analytical equations for  the doublon dispersion [Sup.~Mat.~I].

New types of two-photon excitations emerge in a finite array, namely, the single-photon edge states when one photon is localized at the edge of the lattice, while the other one moves along the lattice, and also the doublon edge states where two bound photons are localized at the edge.  
The single-photon edge states exist only  at the edge with weak tunneling link and have the energy $\eps=\pm\sqrt{J_1^2+J_2^2+2J_1J_2\cos 2k}$, where $-\pi/2<k<\pi/2$ is a real wave number of the  delocalized photon. Quite surprisingly,  contrary to the case when  $J_1=J_2$, the internal and center-of-mass degrees of freedom of doublons are mixed at the edge [Sup.~Mat.~VI]. We have the dimerization brings massive difficulties into the application of the Bethe ansatz  and instead we resorted to the numerical diagonalization of the Hamiltonian in the finite system.


The   results for an array with $N=31$ cavities with $J_2=5J_1$ and a given interaction strength $U=8J_1$  are shown in Fig.~\ref{fig:Distr}. Panel (a) presents the bands of quasi-independent photons (green), one-photon edge states (blue), bulk doublons (red). 
The first Brillouin zone for doublons corresponds to the range $-\pi/2<k<\pi/2$, while $-\pi<k<\pi$ for quasi-independent photons. Horizontal lines show doublon edge states localized at the edge with strong tunneling link (black) and weak link (purple). The panels (c-f) show the 2D color maps of  the corresponding wavefunctions $|\beta_{nm}|^2$. Doublon localization at the edge with strong link for the uppermost band of even symmetry is in full agreement with our analysis above.  We also observe doublon  states localized at the opposite edge with the weak link that are facilitated by the interaction with the single-photon edge states.

A quite interesting phenomenon is the decay of both bulk and edge doublon states due to their interaction with the  quasi-independent photons and the single-photon edge states, respectively.
There exist four bulk doublon bands in total, two of which are unstable against the decay into a pair of quasi-independent photons. We refer to this situation as {\it doublon collapse}: no bound solution with complex $\kappa$ can be found for a certain range of $k$.
 For instance, the third from top doublon band in Fig.~\ref{fig:Distr}(a) is unstable and exists only for the wave vectors close to $\pm \pi/2$. Such effect is  illustrated in Fig.~\ref{fig:Distr}(b) in a larger scale. Additional analysis of doublon collapse is provided in Supplemental Materials, Sec.~II.

\begin{figure}[t]
    \begin{center}
    \includegraphics[width=0.95\linewidth]{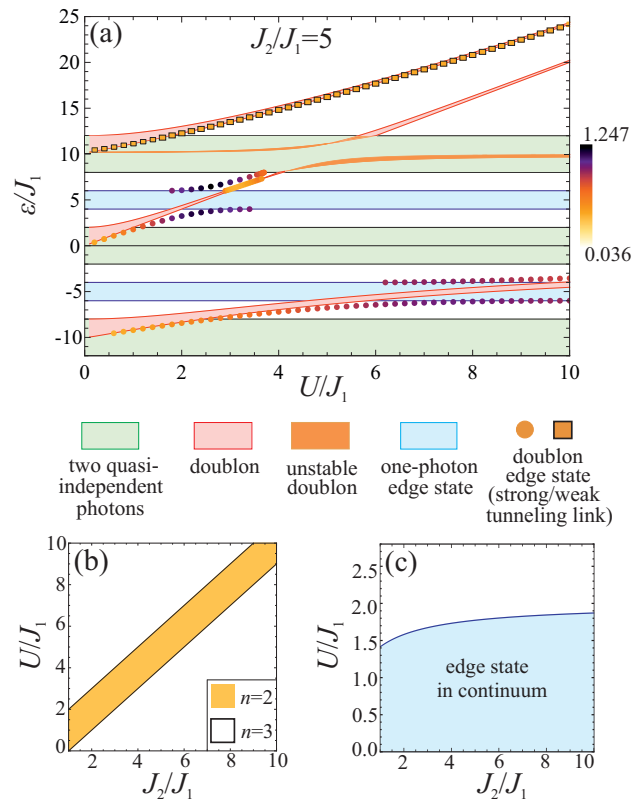}
    \caption{(a) Dependence of the energy bands of various two-photon excitations  on the interaction strength.
    The calculation has been performed for  $J_2/J_1=5$. Color bars show the degree of doublon edge state localization. (b) The number of stable bulk doublon branches as a function of tunneling constants ratio $J_2/J_1$ and nonlinearity parameter $U/J_1$. (c) Phase diagram  for  the  doublon state  in the continuum, localized at the edge with greater tunneling constant. }
    \label{fig:PhaseD}
    \end{center}
    \end{figure}

The stability regions for the bulk and edge states can be further traced by their energy dependence  on the interaction strength, shown in Fig.~\ref{fig:PhaseD}(a).  Color of the circles and squares depicting doublon edge states in Fig.~\ref{fig:PhaseD}(a) encodes their degree of localization $\alpha$ defined as $\alpha=\lim_{n\rightarrow\infty}\ln\left|\beta_{11}/\beta_{n+1,n+1}\right|/n$. The number of stable bulk
doublon bands varies from 2 to 3 depending on the value of $U$, see Fig.~\ref{fig:PhaseD}(b).
The uppermost band with $\eps\approx \eps^{(+)}=2U+J_1^2/U$ and the corresponding edge state, localized at the strong link edge, [squares in
Fig.~\ref{fig:PhaseD}(a)], always remain stable. For weak nonlinearity $U\lesssim 2J_1$ this edge state exists in the continuum of quasi-independent photons [Fig.~\ref{fig:PhaseD}(c)]. 
We have numerically verified that it retains exponential localization in the continuum [Sup.~Mat.~III] and that it exists for an arbitrary ratio of the tunneling constants $J_2/J_1\ne 1$.
 The second from top doublon band, with $\eps_-\approx 2U$ does not have edge states.
The states, localized at the weak link edge [circles in Fig.~\ref{fig:PhaseD}(a,b)], can collapse or revive as function of $U$ due to the interaction with  one-photon edge states (blue bands).

To summarize, our  condition for the edge states of bound photon pairs in terms of the local vertex connectivity of the quantum walks graph   $\kappa>4$ has been fully confirmed by the numerical calculation. The predicted edge state of topmost even photon pair band is localized at the edge with greater tunneling constant, where the single-photon edge states are absent. This state exists for all interaction strengths and for arbitrary unequal  tunneling constants. Its  stability  against the interaction with the continuum of quasi-independent photons might be a sign of   topological protection~\cite{Zhen2014}. It remains to be understood if any meaningful topological invariant aside from the local vertex connectivity can be assigned to the doublon bands. We are not aware of  approaches that
can handle the composite quasiparticle decay, inherently present in the problem.
 For instance, the Zak phase, typically used as an evidence of the topological character of the noninteracting Su-Schrieffer-Heeger model~\cite{Shen}, is not informative here. Our calculations [Sup.~Mat.~VII] show that the Zak phase is equal to $\pi$, if $J_2>J_1$, and it is  equal to $0$, when $J_2<J_1$. This result is the same for all the four doublon bands provided that they are stable and hence does not  explain why  only the state from the  topmost band can be localized at the edge with greater tunneling constant. The breakdown of Bethe anzatz at the edge might further  hinder the application of traditional bulk-boundary correspondence.
 On the contrary, the  vertex connectivity of the quantum walks can be analyzed for various  composite particles (e.g. photon triplets \cite{Flach}) in complex 1D lattices.

Thus, the simple appearance of the two-particle Su-Schrieffer-Heeger model is deceptive. It  uncovers a wide spectrum of fundamental phenomena such as  interaction-induced edge localization and decay of the bound bulk and edge quasiparticles into the weakly interacting ones. Our results may be useful for a whole range of quantum systems  described by Bose-Hubbard model thus paving a way to nonlinear topological physics. The simplest demonstration  might be provided by   waveguide lattices where  the quantum walks of noninteracting \cite{Peruzzo2010,Schreiber2012,Solntsev2014,Hafezi2016}  and interacting photons ~\cite{Corrielli2013} can be emulated classically.

The authors acknowledge valuable discussions with D.~Yudin, G.~Zhilin,  M. Hafezi, I.S.~Sinev, A.K.~Samusev,  A.V. Poshakinskiy, M.M.~Glazov, A.A.~Sukhorukov and Yu.S.~Kivshar. 
This work was supported by the ``Dynasty'' foundation, investigation of bulk doublon dispersion was supported by the Russian Foundation for Basic Research (Grant No.~15-32-20866), calculation of Zak phase was supported by the Russian Science Foundation (Grant No.~16-19-10538).  ANP was  supported by the Russian President Grant No. MK-8500.2016.2.

%

\setcounter{figure}{0}
\setcounter{equation}{0}
\renewcommand{\thefigure}{S\arabic{figure}}
\renewcommand{\theequation}{S\arabic{equation}}

\onecolumngrid
\appendix 
\large
\begin{center}
{\bf Supplementary Materials}
\end{center}
\section{\large I. Dispersion of two-photon excitations in an infinite array}\label{sec:Dispersion}
Since the system Hamiltonian [Eq.~\eqref{eq:Hamiltonian} in the main text] commutes with the operator $\hat{N}=\sum\limits_{m}\,\num{m}$, the total number of photons is conserved and the two-photon wave function can be represented as
\begin{equation}\label{TwoPhotWave}
\bra{\psi}=\sum\limits_{n}\,\sqrt{2}\,\beta_{nn}\,\bra{2_n}+\sum\limits_{m\not=n}\,\beta_{mn}\,\bra{1_m 1_n}\:,
\end{equation}
where $\bra{2_n}\equiv\bra{1_n 1_n}$, and $\bra{1_m 1_n}$ denotes the stationary state when one photon is located in $m$-th cavity, and the other one is located in the $n$-th cavity. Without a loss of generality we assume that $\beta_{mn}=\beta_{nm}$.

\begin{figure}[b]
\begin{center}
\includegraphics[width=0.4\linewidth]{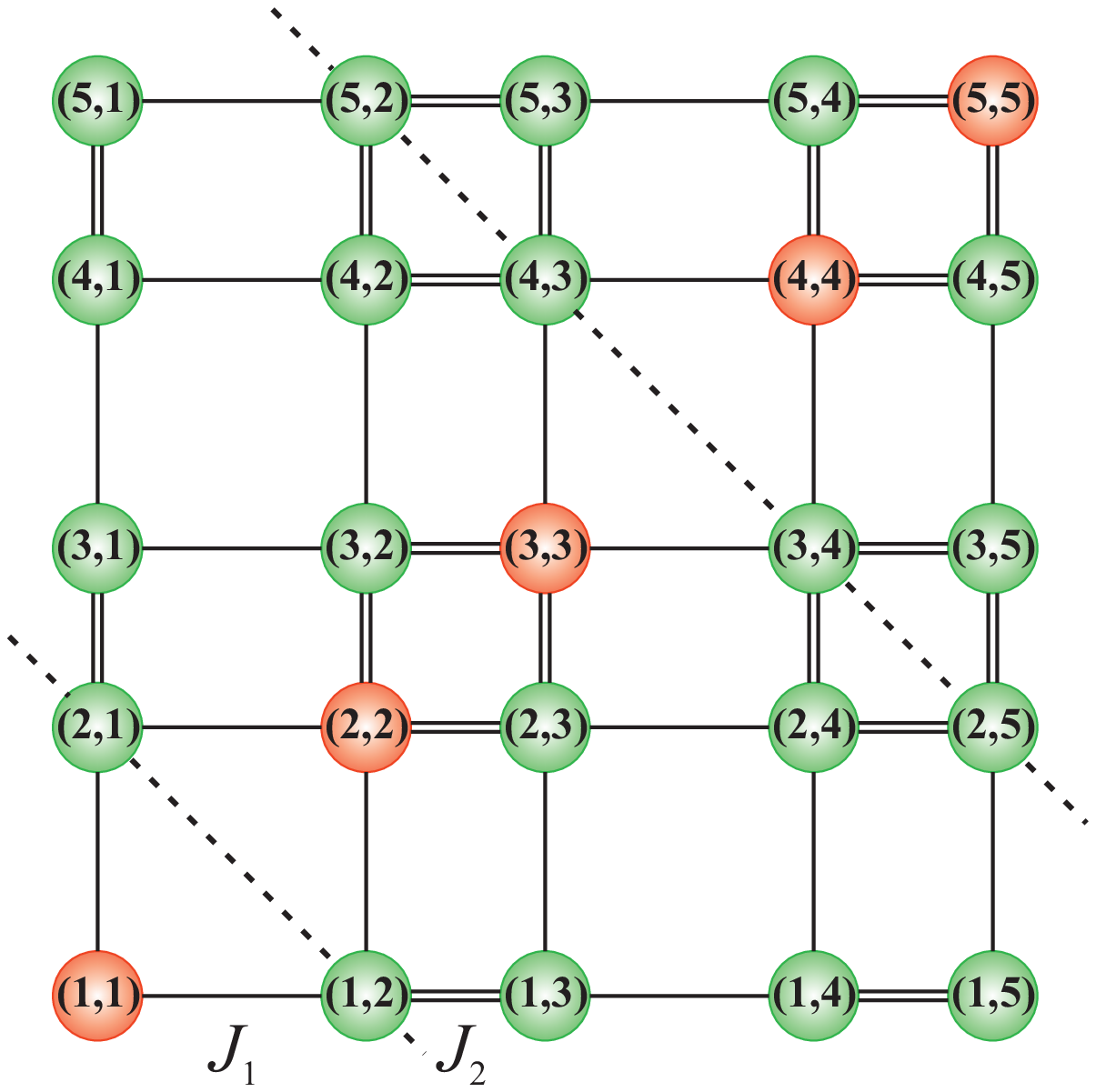}
\caption{Mapping of the one-dimensional two-photon problem onto the two-dimensional single-particle problem. Dashed lines indicate the boundaries of the unit cell chosen in the infinite array to calculate the Zak phase for doublons.}
\label{ris:TwoDimensional}
\end{center}
\end{figure}

Substituting this wave function and the system Hamiltonian $\hat{H}$ into the eigenvalue equation
\begin{equation}\label{Eig}
\hat{H}\,\bra{\psi}=\left(\eps+2\,\omega_0\right)\,\bra{\psi}
\end{equation}
we obtain the system of linear equations with respect to the expansion coefficients $\beta_{mn}$:
{\begin{align}
& (\eps-2\,U)\,\beta_{2m,2m}=-2\,J_1\,\beta_{2m-1,2m}-2\,J_2\,\beta_{2m,2m+1}\:,\label{B1}\\
& (\eps-2\,U)\,\beta_{2m+1,2m+1}=-2\,J_1\,\beta_{2m+1,2m+2}-2\,J_2\,\beta_{2m,2m+1}\:,\nonumber\\
& \eps\,\beta_{2m,2m+2n}=-J_1\,\left(\beta_{2m-1,2m+2n}+\beta_{2m,2m+2n-1}\right)-J_2\,\left(\beta_{2m+1,2m+2n}+\beta_{2m,2m+2n+1}\right)\:,\nonumber\\
& \eps\,\beta_{2m,2m+2n-1}=-J_1\,\left(\beta_{2m-1,2m+2n-1}+\beta_{2m,2m+2n}\right)-J_2\,\left(\beta_{2m+1,2m+2n-1}+\beta_{2m,2m+2n-2}\right)\:,\nonumber\\
& \eps\,\beta_{2m+1,2m+2n}=-J_1\,\left(\beta_{2m+2,2m+2n}+\beta_{2m+1,2m+2n-1}\right)-J_2\,\left(\beta_{2m,2m+2n}+\beta_{2m+1,2m+2n+1}\right)\:,\nonumber\\
& \eps\,\beta_{2m-1,2m-1+2n}=-J_1\,\left(\beta_{2m,2m-1+2n}+\beta_{2m-1,2m+2n}\right)-J_2\,\left(\beta_{2m-2,2m-1+2n}+\beta_{2m-1,2m-2+2n}\right)\:.\nonumber
\end{align}}
In Eqs.~\eqref{B1} one has $n\geq 1$. The system of equations can be interpreted as a two-dimensional problem for a single particle. This equivalent two-dimensional problem is illustrated in Fig.~\ref{ris:TwoDimensional}.

The expansion coefficients $\beta_{mn}$ can be searched in the form of the standard Bethe ansatz~\cite{Essler}:
\begin{equation}\label{Bethe1}
\beta_{mn}=C_{j(m,n)}\,e^{i\,k_1\,m+i\,k_2\,n}\:,
\end{equation}
where $j(m,n)=1,2,3,4$ for the even-even, even-odd, odd-even and odd-odd pairs of indices $(m,n)$, respectively. We assume that $m\leq n$ in Eq.~\eqref{Bethe1}. In the case when $m>n$, the relation $\beta_{mn}=\beta_{nm}$ is used.

Analysis of Eqs.~\eqref{B1} with the ansatz Eq.~\eqref{Bethe1} yields the law of dispersion of two-photon excitations
\begin{equation}\label{Disp}
\begin{split}
\eps^4-4\,\eps^2\,\left[J_1^2+J_2^2+2J_1J_2\,\cos (k_1+k_2)\,\cos (k_1-k_2)\right]\\+16\,J_1^2\,J_2^2\,\sin^2 (k_1+k_2)\,\sin^2 (k_1-k_2)=0\:.
\end{split}
\end{equation}
Importantly, Eq.~\eqref{Disp} describes all considered types of two-photon excitations: bound pairs and quasi-independent pairs,  bulk pairs and edge pairs. Equation~\eqref{Disp} can be rearranged in the equivalent form
\begin{equation}\label{Disp1}
\eps=\pm\sqrt{J_1^2+J_2^2+2J_1J_2\,\cos 2k_1}\pm\sqrt{J_1^2+J_2^2+2J_1J_2\,\cos 2k_2}\:.
\end{equation}
In the case of real $k_1$ and $k_2$ this equation describes the states of quasi-independent photons. The energy of such state is represented as a sum of single-photon energies. Both Bloch wave numbers $k_1$ and $k_2$ are varying in the range from $-\pi/2$ to $\pi/2$.

In addition to the quasi-independent photon states,  the photon-photon interactions determined by the $\propto U$ term of the Bose-Hubbard model give rise to the other type of two-photon states, namely, bound photon pairs (doublons)~\refPR{Winkler}. Such excitations are characterized by complex $k_1$ and $k_2$ of the form $k_1=(k-\varkappa)/2$, $k_2=(k+\varkappa)/2$, where $k$ is a real number describing the motion of the photon pair as a whole, while a complex number $\varkappa$ describes the relative motion of photons. The latter number is assumed to have a positive imaginary part, $\Im \varkappa>0$, that captures the effect of photons binding. We note  that the first Brillouin zone for doublon is one-dimensional and spans from $-\pi/2$ to $\pi/2$.

To obtain the ansatz in the form applicable to doublons, we rewrite Eq.~\eqref{Disp} as follows:
\begin{equation}\label{Disp2}
2\,J_1^2\,J_2^2\,\sin^2 k\cdot\xi^2+\eps^2\,J_1\,J_2\,\cos k\cdot\xi-
\frac{1}{8}\,\left[\eps^4-4\,\eps^2\,\left(J_1^2+J_2^2\right)+16\,J_1^2\,J_2^2\,\sin^2 k\right]=0\:,
\end{equation}
where $\xi=\cos\kap$. Importantly, for the fixed values of doublon energy $\eps$ and Bloch wave number $k$ Eq.~\eqref{Disp2} is a quadratic equation with respect to $\xi$. This means, that in the general case there exist \textit{two} values of $\kap$ with the positive imaginary part corresponding to the given Bloch wave number $k$ and energy $\eps$; these two roots are denoted further as $\kap$ and $\bar{\kap}$. Therefore, to describe the dispersion of doublons, we use a modified Bethe ansatz:
\begin{equation}\label{Bethe2}
\beta_{mn}=e^{i\,k/2\,(m+n)}\,\left[C_{j(m,n)}\,e^{i\kap/2\,(n-m)}+\bar{C}_{j(m,n)}\,e^{i\bar{\kap}/2\,(n-m)}\right]
\end{equation}
with $j(m,n)=1,2,3,4$ for even-even, even-odd, odd-even and odd-odd pairs of indices $(m,n)$ and $m\leq n$. This ansatz yields the following system of linear equations:
\begin{align}
& \eps\,C_1+\left(J_1\,e^{-i k_2}+J_2\,e^{i k_2}\right)\,C_2+\left(J_1\,e^{-i k_1}+J_2\,e^{i k_1}\right)\,C_3=0\:,\label{C1}\\
& \left(J_1\,e^{i k_2}+J_2\,e^{-i k_2}\right)\,C_1+\eps\,C_2+\left(J_1\,e^{-i k_1}+J_2\,e^{i k_1}\right)\,C_4=0\:,\label{C2}\\
& \left(J_1\,e^{i k_1}+J_2\,e^{-i k_1}\right)\,C_1+\eps\,C_3+\left(J_1\,e^{-i k_2}+J_2\,e^{i k_2}\right)\,C_4=0\:,\label{C3}\\
& \left(J_1\,e^{i k_1}+J_2\,e^{-i k_1}\right)\,C_2+\left(J_1\,e^{i k_2}+J_2\,e^{-i k_2}\right)\,C_3+\eps\,C_4=0\:,\label{C4}\\
& \eps\,\bar{C}_1+\left(J_1\,e^{-i \bar{k}_2}+J_2\,e^{i \bar{k}_2}\right)\,\bar{C}_2+\left(J_1\,e^{-i \bar{k}_1}+J_2\,e^{i \bar{k}_1}\right)\,\bar{C}_3=0\:,\label{D1}\\
& \left(J_1\,e^{i \bar{k}_2}+J_2\,e^{-i \bar{k}_2}\right)\,\bar{C}_1+\eps\,\bar{C}_2+\left(J_1\,e^{-i \bar{k}_1}+J_2\,e^{i \bar{k}_1}\right)\,\bar{C}_4=0\:,\label{D2}\\
& \left(J_1\,e^{i \bar{k}_1}+J_2\,e^{-i \bar{k}_1}\right)\,\bar{C}_1+\eps\,\bar{C}_3+\left(J_1\,e^{-i \bar{k}_2}+J_2\,e^{i \bar{k}_2}\right)\,\bar{C}_4=0\:,\label{D3}\\
& \left(J_1\,e^{i \bar{k}_1}+J_2\,e^{-i \bar{k}_1}\right)\,\bar{C}_2+\left(J_1\,e^{i \bar{k}_2}+J_2\,e^{-i \bar{k}_2}\right)\,\bar{C}_3+\eps\,\bar{C}_4=0\:,\label{D4}\\
& (\eps-2\,U)\,\left(C_1+\bar{C}_1\right)=-2 J_1\,\left(C_3\,e^{-i k_1}+\bar{C}_3\,e^{-i \bar{k}_1}\right)-2 J_2\,\left(C_2\,e^{i k_2}+\bar{C}_2\,e^{i\bar{k}_2}\right)\:,\label{CD1}\\
& (\eps-2\,U)\,\left(C_4+\bar{C}_4\right)=-2 J_1\,\left(C_3\,e^{i k_2}+\bar{C}_3\,e^{i \bar{k}_2}\right)-2 J_2\,\left(C_2\,e^{-i k_1}+\bar{C}_2\,e^{-i \bar{k}_1}\right)\:.\label{CD2}
\end{align}
Here, $k_{1,2}=(k\mp\kap)/2$ and $\bar{k}_{1,2}=(k\mp\bar{\kap})/2$. Note that the determinant of Eqs.~\eqref{C1}-\eqref{C4} as well as Eqs.~\eqref{D1}-\eqref{D4} is zero due to the same dispersion equation Eq.~\eqref{Disp2}. Thus, the calculation of the doublon dispersion can be accomplished as follows:
\begin{itemize}
\item Express $\kap$ and $\bar{\kap}$ as functions of energy $\eps$ and wave number $k$ from Eq.~\eqref{Disp2}.
\item Express $C_2$, $C_3$ and $C_4$ via $C_1$, $\eps$ and $k$ from Eqs.~\eqref{C1}-\eqref{C4}.
\item Express $\bar{C}_2$, $\bar{C}_3$ and $\bar{C}_4$ via $\bar{C}_1$, $\eps$ and $k$ from Eqs.~\eqref{D1}-\eqref{D4}.
\item Obtain a system of homogeneous linear equations with the unknown variables $C_1$ and $\bar{C}_1$ from  Eqs.~\eqref{CD1}, \eqref{CD2}. The coefficients in the obtained system depend on the doublon energy $\eps$ and the wave number $k$.
\item Doublon  dispersion equation is deduced setting to zero the determinant of the derived system of equations.
\end{itemize}

\section{\large II. Doublon collapse}\label{sec:Collapse}
\begin{figure}[b]
\begin{center}
\includegraphics[width=0.8\linewidth]{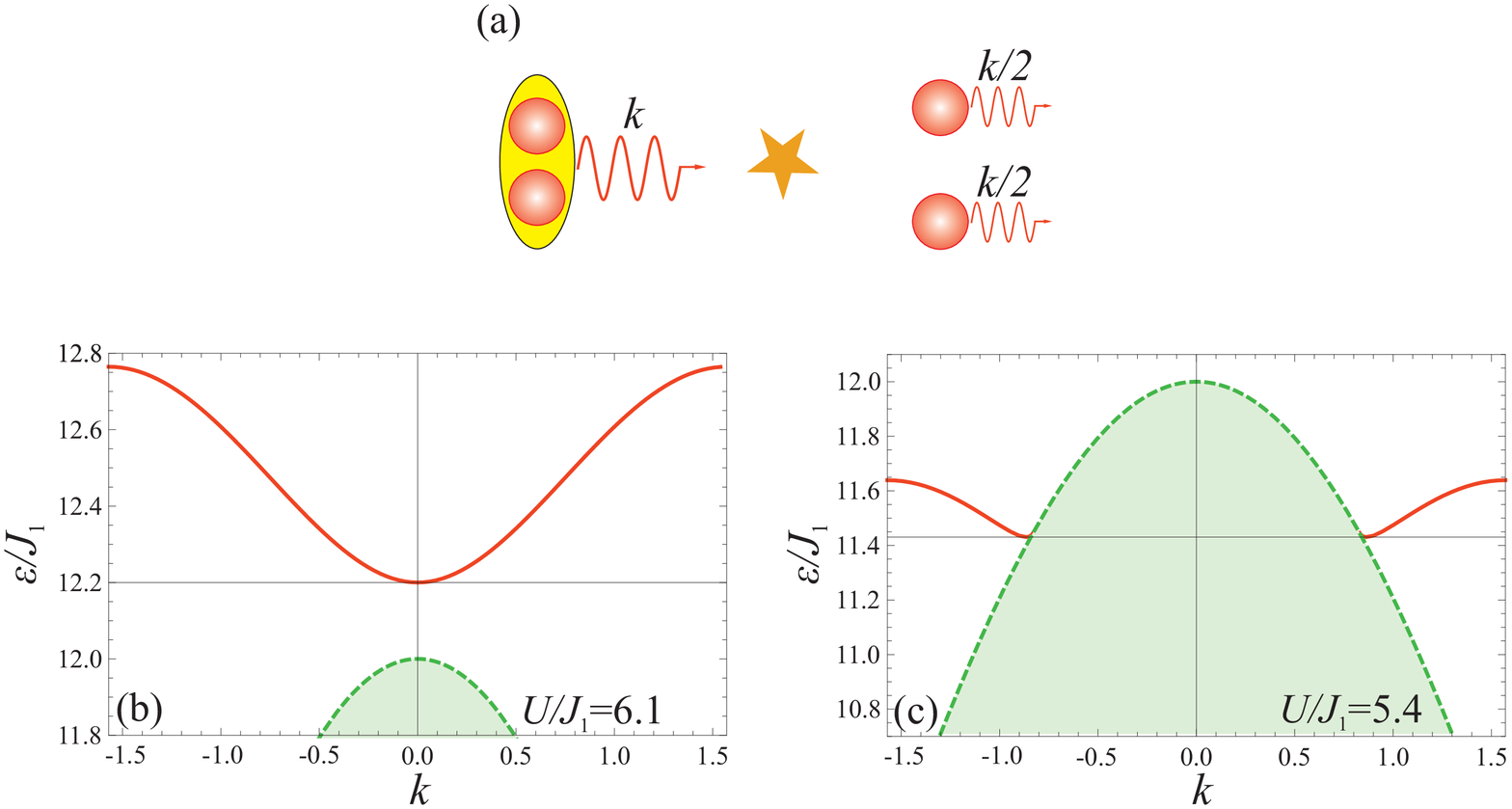}
\caption{(a) Illustration of the doublon collapse effect: the bound photon pair decays into a pair of quasi-independent photons. (b) Dispersion for $J_2/J_1=5$, $U/J_1=6.1$: doublon dispersion curve (red) almost touches the quasi-independent photons band (green). (c) $J_2/J_1=5$, $U/J_1=5.4$: doublon collapses for sufficiently small wave numbers and exists only in the range of $k$ close to the boundary of the first Brillouin zone.}
\label{ris:DoublonCollapse}
\end{center}
\end{figure}

We analyze the dispersion equation Eq.~\eqref{Disp2} in the special case $k=0$ denoting $\eps(k=0)$ by $\eps_0$. It is straightforward to show that
\begin{equation}\label{Collapse1}
\cos\kap_0=\frac{\eps_0^2-4\,\left(J_1^2+J_2^2\right)}{8J_1J_2}\:.
\end{equation}
Photons will be bound together if Im~$\kap_0>0$, i.e. $\left|\cos\kap_0\right|>1$. In the opposite scenario $\left|\cos\kap_0\right|<1$, $\kap_0$ is purely real, photons are no longer confined to each other and doublon thus decays into the pair of quasi-independent photons. Further we refer to this phenomenon as doublon collapse. It can be shown  that the condition for doublon collapse reads
\begin{equation}\label{Collapse2}
2\,\left|J_2-J_1\right|<|\eps_0|<2\,\left|J_2+J_1\right|\:.
\end{equation}
On the other hand, the energy intervals $[-2(J_2+J_1);-2(J_2-J_1)]$, $[-2J_1;2J_1]$ and $[2(J_2-J_1);2(J_2+J_1)]$ for $J_2>J_1$ correspond to the energy bands of quasi-independent photons. Thus, we conclude that if the doublon energy $\eps_0$ falls into the upper or lower energy band of quasi-independent photons, the doublon becomes unstable and collapses. Besides that, our calculations show that the doublon is always stable for the values of $k$ close to $\pm\pi/2$. The phenomenon of doublon collapse is illustrated in more detail in Fig.~\ref{ris:DoublonCollapse} where two panels correspond to different strengths of interaction.

\section{\large III. Two-photon states in a semi-infinite array}\label{sec:Edge}
New types of two-photon excitations emerge, if the array has an edge. In this section, we analyze various types of edge states in a semi-infinite geometry, when the structure starts from the cavity with the number $n=1$. In such situation system of equations Eq.~\eqref{B1} should be supplemented by the boundary conditions of the form:
\begin{gather}
\left(\eps-2U\right)\,\beta_{11}=-2J_1\,\beta_{12}\:,\label{Bound1}\\
\eps\,\beta_{1,2n}=-J_1\,\left(\beta_{1,2n-1}+\beta_{2,2n}\right)-J_2\,\beta_{1,2n+1}\:,\label{Bound2}\\
\eps\,\beta_{1,2n+1}=-J_1\,\left(\beta_{2,2n+1}+\beta_{1,2n+2}\right)-J_2\,\beta_{1,2n}\:,\label{Bound3}
\end{gather}
where $n\geq 1$. 

First we consider a situation when one photon is localized at the edge, while the other one moves freely along the lattice. We assume ansatz Eq.~\eqref{Bethe1} for $\beta_{mn}$ coefficients. Then Eqs.~\eqref{B1} are compatible with Eqs.~\eqref{Bound1}--\eqref{Bound3} only under the condition $C_1=C_2=0$. Simple analysis reveals that edge states exist if $J_1<J_2$. The energy of the edge state and its localization parameter read
\begin{gather}
\eps=\pm\sqrt{J_1^2+J_2^2+2J_1J_2\,\cos 2k_2}\label{Edge1}\:,\\
e^{2ik_1}=-J_1/J_2\:.\label{Edge2}
\end{gather}
In fact, the edge state described by Eqs.~\eqref{Edge1}-\eqref{Edge2} is a direct analog of that existing in a one-dimensional single-particle Su-Schrieffer-Heeger model.

\begin{figure}[t]
\begin{center}
\includegraphics[width=0.7\linewidth]{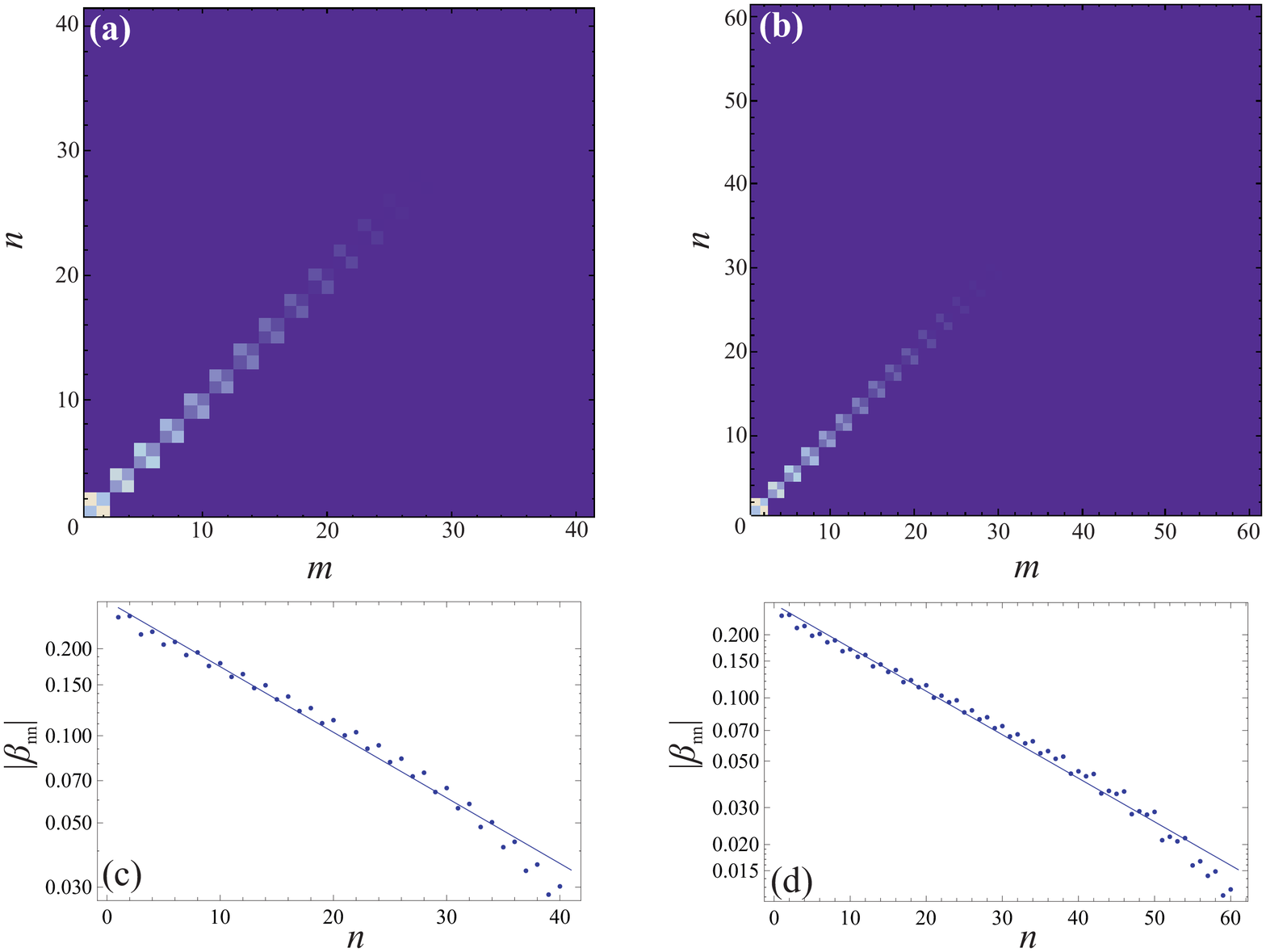}
\caption{Numerical investigtion of the doublon edge state in continuum for $J_1/J_2=5$, $U=1.0$. (a,b) Calculated probability distributions for a finite array composed of (a) 41, (b) 61 cavities. (c,d) Dependence of $|\beta_{nn}|$ on $n$ in a logarithmic scale for a finite array composed of (c) 41, (d) 61 cavities. Exponential localization is clearly observed.}
\label{ris:EdgeInContinuum}
\end{center}
\end{figure}

Another type of edge modes is represented by the edge states of bound photon pairs. However, their analytical description is extremely cumbersome (if possible at all, see Sec. VI), and therefore we resorted to the full numerical diagonalization of the finite system Hamiltonian. In particular, we demonstrate that doublon edge states can arise in continuum. The results of numerical diagonalization of the system Hamiltonian are presented in Fig.~\ref{ris:EdgeInContinuum}.

\begin{figure}[ht]
\begin{center}
\includegraphics[width=0.9\linewidth]{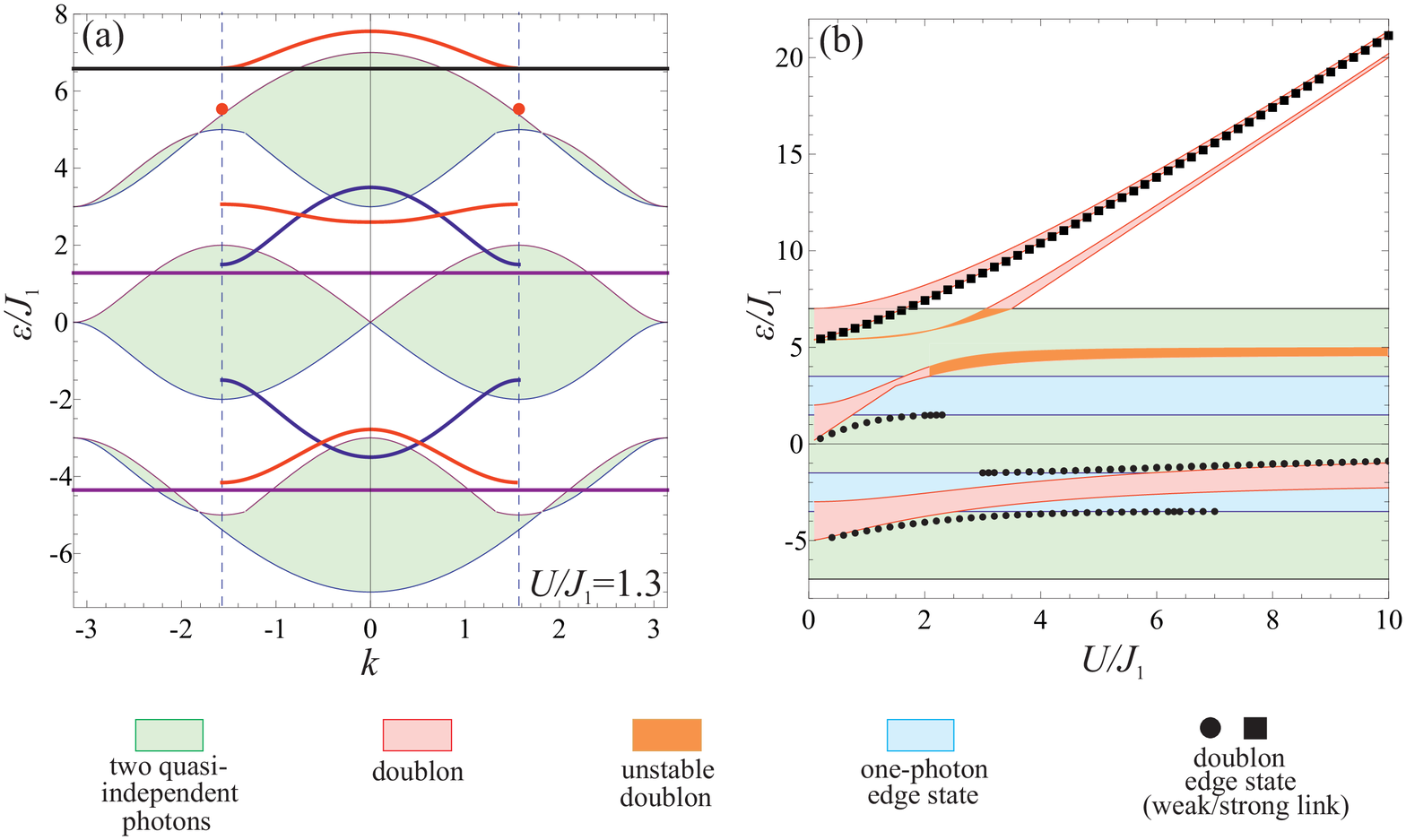}
\caption{(a) Dispersion of various types of two-photon excitations in a semi-infinite array for fixed tunneling constants ratio $J_2/J_1=2.5$ and fixed interaction strength $U/J_1=1.3$. Red dots indicate collapse of the second from top doublon band. Black and purple horizontal lines indicate the doublon states localized at the edge with strong and weak tunneling links, respectively. (b) Energy bands of various types of two-photon excitations versus interaction strength $U/J_1$.}
\label{ris:Diagram-J-25}
\end{center}
\end{figure}

In order to provide more insight into the properties of different two-photon states, we plot the dispersion of various types of two-photon excitations for a fixed tunneling constants ratio $J_2/J_1=2.5$ and a fixed interaction strength $U/J_1=1.3$ (Fig.~\ref{ris:Diagram-J-25}a, cf. with Fig.~\ref{fig:PhaseD} in the main text). Such  values of parameters correspond to the partial overlap of quasi-independent photons energy band with the energy band of single-photon edge states. Therefore, some types of doublon edge states present for larger tunneling constants ratio, e.g. for $J_2/J_1=5$, disappear. This tendency is further illustrated by Fig.~\ref{ris:Diagram-J-25}b. Figure~\ref{ris:Diagram-J-25}b shows also that only two from four doublon energy bands are strongly affected by the nonlinearity, exhibiting almost linear growth with the interaction strength $U$. The reason for such behavior is that photons are predominantly localized in the same cavity for two upper doublon bands, while for two lower doublon bands  photons are mainly localized in the neighboring cavities.

\section{\large IV. Effective Hamiltonian for two upper doublon bands}\label{sec:Heff}
As it is pointed out in the previous section, energies of the two from four doublon bands depend almost linearly on $U$. Thus, in the limit of strong interaction $U\gg \text{max}\left(J_1,J_2\right)$ these two bands are well separated from the remaining states (see Fig.~\ref{ris:Diagram-J-25}b) and the two photons are well confined to each other. Therefore, it is possible to develop an  effective one-dimensional model describing the properties of the upper doublon bands.

Examining the calculated probability distributions for the two upper doublon bands, we notice that the $\beta_{mn}$ coefficients with $|m-n|\geq 3$ are always negligible. Therefore, for this special case the system of equations Eqs.~\eqref{B1} together with the boundary conditions Eqs.~\eqref{Bound1}-\eqref{Bound3} can be truncated. Excluding from these truncated equations $\beta_{n,n+1}$, $\beta_{n,n+2}$ and leaving only the terms $\alpha_n\equiv\beta_{nn}$, we obtain:
\begin{gather}
\eps\,\alpha_1=(2U+j_1+\tau)\,\alpha_1+(j_1+2\tau)\,\alpha_2+\tau\,\alpha_3\:,\label{Al1}\\
\eps\,\alpha_2=(2U+j_1+j_2+5\tau)\,\alpha_2+(j_1+2\tau)\,\alpha_1+(j_2+4\tau)\,\alpha_3+\tau\,\alpha_4\:,\nonumber
\\
\eps\,\alpha_3=(2U+j_1+j_2+6\tau)\,\alpha_3+(j_2+4\tau)\,\alpha_2+(j_1+4\tau)\,\alpha_4+\tau\,\alpha_1+\tau\,\alpha_5\:,\nonumber
\\
\eps\,\alpha_4=(2U+j_1+j_2+6\tau)\,\alpha_4+(j_1+4\tau)\,\alpha_3+(j_2+4\tau)\,\alpha_5+\tau\,\alpha_2+\tau\,\alpha_6\:,
\notag 
\\
\dots\notag
\end{gather}
where $j_{1,2}=2J_{1,2}^2/\eps$ and $\tau=2J_1^2J_2^2/\eps^3$. Equations \eqref{Al1} describe a one-dimensional  lattice with the next-nearest neighbor hopping.

In order to further  simplify the analysis we consider the limit of strong interaction and strongly different tunneling constants $U\gg J_1\gg J_2$. In such case, in the zero-order approximation the array consists of isolated dimers and its eigenstates are
\begin{equation}\label{Eigenstate}
u^{(\pm)}=
\begin{pmatrix}
1/\sqrt{2}\\
\pm 1/\sqrt{2}
\end{pmatrix}\:.
\end{equation}
We rewrite Eqs.~\eqref{Al1} in terms of the vectors $u_m=(\alpha_{2m-1},\alpha_{2m})^T$ and project these equations onto the eigenvectors $u^{(\pm)}$. As a result, we derive the following equations for the projections $\psi_m=\left<\left.u^{(\pm)}\right|u_m\right>$:
\begin{align}
\eps\,\psi_1&=(\spm{\eps_0}+\delta_{\pm}/2)\,\psi_1+t_{\pm}\,\psi_2\:,\label{Psi1}\\
\eps\,\psi_m&=(\spm{\eps_0}+\delta_{\pm})\,\psi_m+t_{\pm}\,\left(\psi_{m-1}+\psi_{m+1}\right)\:,\label{Psi2}
\end{align}
where $\spm{\eps_0}=2U+j_1\pm j_1$, $\delta_{\pm}=j_2+6\,\tau\pm 4\tau$ and $t_{\pm}=\pm j_2/2\pm 2\tau+\tau$. The ``$-$" sign choice corresponds to the antisymmetric eigenstates and describes the second from top doublon energy band. On the contrary, the ``$+$" sign choice corresponds to the symmetric eigenstates and describes the upper doublon energy band. For both signs, the equations Eq.~\eqref{Psi1} and Eq.~\eqref{Psi2} formally correspond to the one-dimensional array with the detuned resonator at the edge. The magnitude of detuning is equal to $\delta_{\pm}/2$. Thus, the well-known condition for edge states emergence reads $\delta_{\pm}-2\left|t_{\pm}\right|>0$ (see Sec. V). From the derived results it is straightforward to show that $\delta_{-}-2|t_{-}|=0$ and $\delta_{+}-2|t_{+}|=4\tau=8J_1^2\,J_2^2/\eps^3\approx J_1^2\,J_2^2/U^3$. Using the correspondence with the one-dimensional  lattice, it is also easy to estimate localization length of doublon edge state associated with the upper band:
\begin{equation}\label{Length}
l=\frac1{\ln\left(\frac{\delta_+}{2\,|t_+|}\right)}\approx \frac{2|t_+|}{\delta_+-2|t_+|}\approx \left(\frac{U}{J_1}\right)^2\:.
\end{equation}
These conclusions are in perfect agreement with the general topological arguments provided in the article main text.

\section{\large V. Edge states in a simple lattice with the detuned edge}\label{sec:SimpleEdge}

In the previous section we have reduced the problem for both symmetric and antisymmetric doublon modes to the following one:
\begin{gather}
\eps\,\psi_1=\Delta\,\psi_1+t\,\psi_2\:,\label{PsiE1}\\
\eps\,\psi_m=t\,(\psi_{m-1}+\psi_{m+1})\:,\mspace{10mu} m\geq 2\:.\label{PsiE2}
\end{gather}
Equations~\eqref{PsiE1},\eqref{PsiE2} correspond formally to the case of one-dimensional semi-infinite lattice with the edge cavity detuning equal to $\Delta$. We search edge states in the system assuming $\psi_m=\psi\,e^{ikm}$ with $k$ having positive imaginary part. It is straightforward to show that $e^{ik}=t/\Delta$. Since $|e^{ik}|<1$ due to nonzero imaginary part of $k$, the condition for edge state existence reads $|\Delta|>|t|$. Localization length of edge state can be calculated as
\begin{equation}\label{LocLengthSim}
l=\frac{1}{\text{Im}\,k}=\frac{1}{\ln|t/\Delta|}\:.
\end{equation}

\section{\large VI. Bethe ansatz breakdown}\label{sec:Breakdown}

In this section we discuss a general ansatz for $\beta_{mn}$ that allows one to investigate analytically all types of two-photon excitations supported by a finite cavity array. Due to the property $\beta_{mn}=\beta_{nm}$, it is sufficient to consider only the coefficients $\beta_{mn}$ with $m\leq n$.

\begin{figure}[b]
\begin{center}
\includegraphics[width=0.45\linewidth]{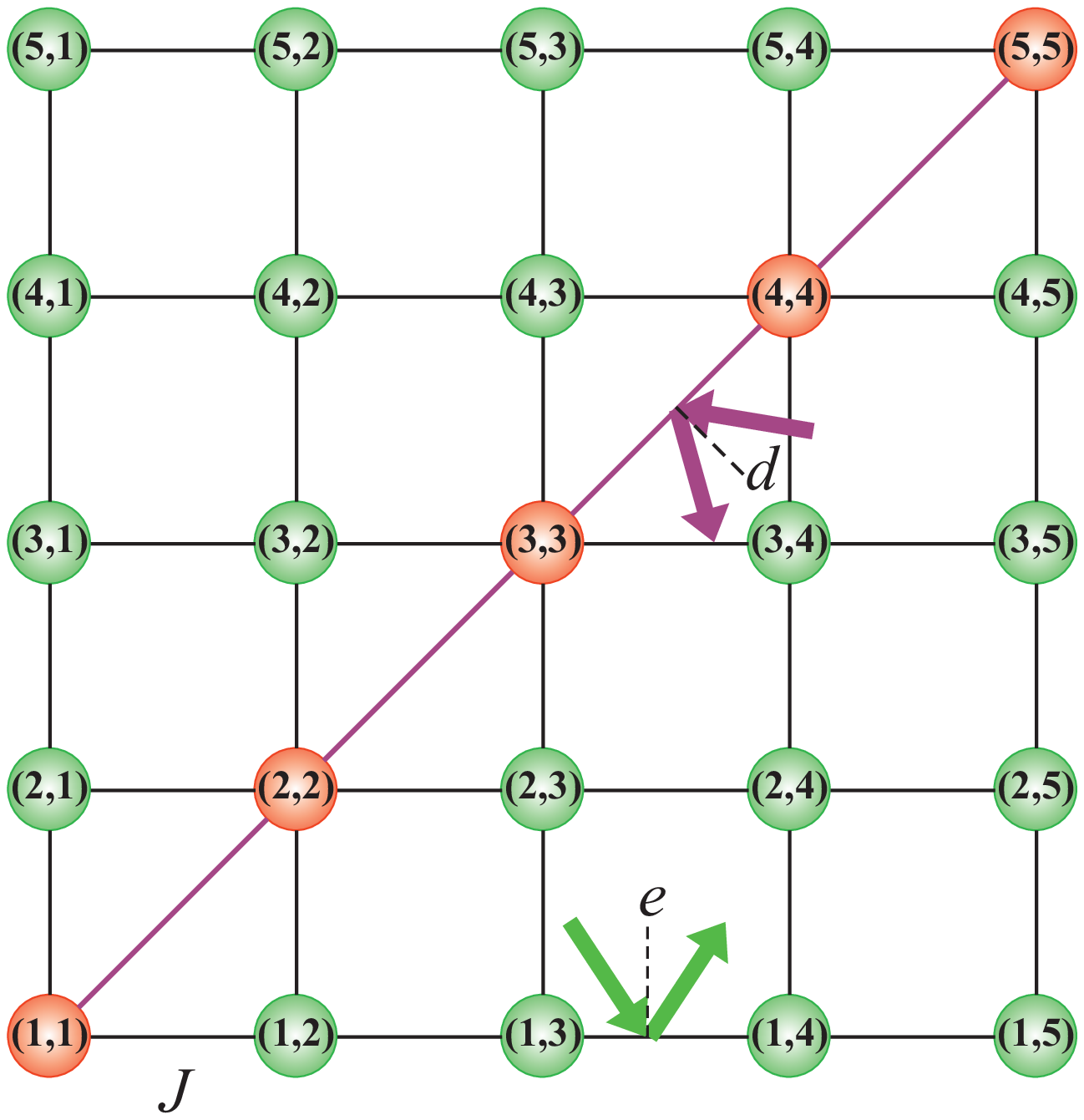}
\caption{Different types of scattering contributing to the Bethe ansatz Eq.~\eqref{FullAnsatz1} in an equivalent two-dimensional problem for a simple array.}
\label{ris:TwoDimScheme}
\end{center}
\end{figure}

To get a general idea of constructing ansatz for $\beta_{mn}$, we consider first a situation with $J_1=J_2$, previously studied in the literature in Refs.~\refPR{Flach,Pinto,Longhi}. We map the two-photon problem onto the corresponding two-dimensional single-particle problem. Two-photon state is represented as a superposition of the wave $C_{j(m,n)}\,e^{ik_1m+ik_2n}$ and various scattered waves which arise from the scattering at the diagonal and at the edge of the two-dimensional sample. Such ``diagonal'' and ``edge'' scattering is described by the operators (Fig.~\ref{ris:TwoDimScheme}).
\begin{gather}
d(k_1,k_2)=(k_2,k_1)\:,\label{DiagScattering}\\
e(k_1,k_2)=(-k_1,k_2)\:.\label{EdgeScattering}
\end{gather}
We note further that the elements $d$ and $e$ describing two types of scattering generate a group of the eighth order, i.e. ansatz for $\beta_{mn}$ will contain eight terms
\begin{equation}\label{FullAnsatz1}
\begin{split}
& \beta_{nm}=C^{(1)}_{j(m,n)}\,e^{ik_1\,n+ik_2\,m}+C^{(2)}_{j(m,n)}\,e^{-ik_1\,n+ik_2\,m}+C^{(3)}_{j(m,n)}\,e^{ik_1\,n-ik_2\,m}+C^{(4)}_{j(m,n)}\,e^{-ik_1\,n-ik_2\,m}+\\
&+C^{(5)}_{j(m,n)}\,e^{ik_2\,n+ik_1\,m}+C^{(6)}_{j(m,n)}\,e^{-ik_2\,n+ik_1\,m}+C^{(7)}_{j(m,n)}\,e^{ik_2\,n-ik_1\,m}+C^{(8)}_{j(m,n)}\,e^{-ik_2\,n-ik_1\,m}\:,
\end{split}
\end{equation}
It is this ansatz that was employed in the analysis of Tamm-Hubbard states in an array with equal tunneling constants $J_1=J_2$~\refPR{Longhi}.

However, the situation becomes much more complicated if $J_1\not=J_2$. Besides the elements $e$ and $d$ describing scattering at the diagonal and the edges, one also needs to take into account the element $x$ that acts on wave vector as follows:
\begin{gather}
x(k_1,k_2)=(\bar{k}_1,\bar{k}_2)\:,\label{XTerm}\\
\bar{k}_1=1/2\,\left(k_1+k_2\right)+1/2\,\text{arccos}\left(-\cos(k_1-k_2)-\frac{\eps^2}{2\,J_1\,J_2}\,\frac{\cos(k_1+k_2)}{\sin^2(k_1+k_2)}\right)\:,\label{BarK1}\\
\bar{k}_2=1/2\,\left(k_1+k_2\right)-1/2\,\text{arccos}\left(-\cos(k_1-k_2)-\frac{\eps^2}{2\,J_1\,J_2}\,\frac{\cos(k_1+k_2)}{\sin^2(k_1+k_2)}\right)\:,\label{BarK2}
\end{gather} 
Indeed, this scattering $x$ is responsible for doublon formation. It turns out that the group generated by the elements $e$, $d$ and $x$ is infinite in the general case. Thus, it is problematic to find the ansatz analogous to Eq.~\ref{FullAnsatz1} and valid for the description of all two-photon states in a finite array. For that reason, we opted to study doublon edge states using full numerical diagonalization of the finite system Hamiltonian.

\section{\large VII. Calculation of the Zak phase for doublons}\label{sec:Zak}
In this section we calculate the Zak phase for doublons propagating in an infinite lattice. To this end, we extract the periodic part $\bra{u_k}$ from the entire doublon wave function calculated in Sec.~I. The choice of the two-photon states comprising the periodic part is illustrated in Fig.~\ref{ris:TwoDimensional}, so $\bra{u_k}$ reads:
\begin{equation}\label{PeriodicPart}
\begin{split}
& \bra{u_k}=\left\lbrace\sqrt{2}\,\beta_{00}\,\bra{2_0}+2\,\sum\limits_{n=1}^{\infty}\,\beta_{-2n,2n}\,\bra{1_{-2n} 1_{2n}}+2\,\sum\limits_{n=1}^{\infty}\,\beta_{2-2n,2n}\,\bra{1_{2-2n} 1_{2n}}\right\rbrace\\+
& \left\lbrace 2\,\sum\limits_{n=1}^{\infty}\,\beta_{2-2n,2n-1}\,\bra{1_{2-2n} 1_{2n-1}}+\sum\limits_{n=1}^{\infty}\,\beta_{-2n,2n-1}\,\bra{1_{-2n} 1_{2n-1}}+\sum\limits_{n=1}^{\infty}\,\beta_{2-2n,2n+1}\,\bra{1_{2-2n} 1_{2n+1}}\right\rbrace\\+
& \left\lbrace 2\,\sum\limits_{n=1}^{\infty}\,\beta_{1-2n,2n}\,\bra{1_{1-2n} 1_{2n}}+\sum\limits_{n=1}^{\infty}\,\beta_{1-2n,2n-2}\,\bra{1_{1-2n} 1_{2n-2}}+\sum\limits_{n=1}^{\infty}\,\beta_{3-2n,2n}\,\bra{1_{3-2n} 1_{2n}}\right\rbrace\\+
& \left\lbrace\sqrt{2}\,\beta_{11}\,\bra{2_1}+2\,\sum\limits_{n=1}^{\infty}\,\beta_{1-2n,2n+1}\,\bra{1_{1-2n} 1_{2n+1}}+2\,\sum\limits_{n=1}^{\infty}\,\beta_{1-2n,2n-1}\,\bra{1_{1-2n} 1_{2n-1}}\right\rbrace\:.
\end{split}
\end{equation}
Note that the chosen unit cell possesses inversion symmetry. Therefore, we expect the Zak phase to be quantized~\refPR{Zak}. Using the modified Bethe ansatz Eq.~\eqref{Bethe2} for the coefficients $\beta_{mn}$, we obtain:
\begin{align}
& \bra{u_k}=\sum\limits_{j=1}^4\,\bra{u_j}\:,\label{P1}\\
& \bra{u_j}=C_j\,\bra{v_j}+\bar{C}_j\,\bra{\bar{v}_j}\:,\label{P2}\\
& \bra{v_1}=\sqrt{2}\,\bra{2_0}+2\,\sum\limits_{n=1}^{\infty}\,e^{2i \kap n}\,\bra{1_{-2n} 1_{2n}}+2\,e^{i k}\,\sum\limits_{n=1}^{\infty}\,e^{i\kap (2n-1)}\,\bra{1_{2-2n} 1_{2n}}\:,\label{P3}\\
& \bra{v_2}=2\,e^{ik/2}\,\sum\limits_{n=1}^{\infty}\,e^{i\kap\,(2n-3/2)}\,\bra{1_{2-2n} 1_{2n-1}}+\notag\\
& e^{-ik/2}\,\sum\limits_{n=1}^{\infty}\,e^{i\kap\,(2n-1/2)}\,\bra{1_{-2n} 1_{2n-1}}+e^{3ik/2}\,\sum\limits_{n=1}^{\infty}\,e^{i\kap\,(2n-1/2)}\,\bra{1_{2-2n} 1_{2n+1}}\:,\label{P4}
\end{align}

\begin{align}
& \bra{v_3}=2\,e^{ik/2}\,\sum\limits_{n=1}^{\infty}\,e^{i\kap(2n-1/2)}\,\bra{1_{1-2n} 1_{2n}}+\notag\\
& e^{-ik/2}\,\sum\limits_{n=1}^{\infty}\,e^{i\kap(2n-3/2)}\,\bra{1_{1-2n} 1_{2n-2}}+e^{3ik/2}\,\sum\limits_{n=1}^{\infty}\,e^{i\kap(2n-3/2)}\,\bra{1_{3-2n} 1_{2n}}\:,\label{P5}\\
& \bra{v_4}=\sqrt{2}\,e^{ik}\,\bra{2_1}+2\,e^{ik}\,\sum\limits_{n=1}^{\infty}\,e^{2i\kap n}\,\bra{1_{1-2n} 1_{2n+1}}+2\,\sum\limits_{n=1}^{\infty}\,e^{i\kap(2n-1)}\,\bra{1_{1-2n} 1_{2n-1}}\:.\label{P6}
\end{align}
Expressions analogous to Eqs.~\eqref{P3}-\eqref{P6} are valid for $\bra{\bar{v}_j}$. The only difference is in the replacement of $\kap$ by $\bar{\kap}$. Berry connection is defined by the equations
\begin{align}
 A(k)=&\sum\limits_{j=1}^4\,A_j(k)\:,\label{A1}\\
 A_j(k)\equiv &i\,\skvk{u_j}{u_j}=i\,C_j^*\,\df{C_j}\,\skvv{v_j}{v_j}+i\,C_j^*\,\df{\bar{C}_j}\,\skvv{v_j}{\bar{v}_j}+i\,\bar{C}_j^*\,\df{C_j}\,\skvv{\bar{v}_j}{v_j}+\notag \\&i\,\bar{C}_j^*\,\df{\bar{C}_j}\,\skvv{\bar{v}_j}{\bar{v}_j}+
i\,C_j^*\,C_j\,\skvk{v_j}{v_j}+i\,C_j^*\,\bar{C}_j\,\skvk{v_j}{\bar{v}_j}\notag\\+&i\,\bar{C}_j^*\,C_j\,\skvk{\bar{v}_j}{v_j}+i\,\bar{C}_j^*\,\bar{C}_j\,\skvk{\bar{v}_j}{\bar{v}_j}\:.\label{A2}
\end{align}
Scalar products in Eq.~\eqref{A2} read:
\begin{align}
& \skvv{v_1}{v_1}=2\,\,\frac{1+x}{1-x}\:,\label{S1}\\
& \skvv{v_2}{v_2}=\sqrt{x}\,\,\frac{4+2 x}{1-x^2}\:,\label{S2}\\
& \skvv{v_3}{v_3}=\sqrt{x}\,\,\frac{4x+2}{1-x^2}\:,\label{S3}\\
& \skvv{v_4}{v_4}=2\,\,\frac{1+x}{1-x}\:,\label{S4}\\
& \skvk{v_1}{v_1}=4i\,\df{\kap}\,\frac{x}{(1-x)^2}+4i\,\frac{x}{1-x^2}\:,\label{S5}\\
& \skvk{v_2}{v_2}=\frac{i\sqrt{x}\,(2+x)}{1-x^2}+i\df{\kap}\,\sqrt{x}\,\,\frac{2+3x+6x^2+x^3}{(1-x^2)^2}\:,\label{S6}\\
& \skvk{v_3}{v_3}=\frac{i\,\sqrt{x} (2x+1)}{1-x^2}+i\df{\kap}\,\sqrt{x}\,\,\frac{1+6x+3x^2+2x^3}{(1-x^2)^2}\:,\label{S7}\\
& \skvk{v_4}{v_4}=2i\,\,\frac{1+x^2}{1-x^2}+4i\,\df{\kap}\,\frac{x}{(1-x)^2}\:.\label{S8}
\end{align}
In these scalar products, $x=\left(e^{i\kap}\right)^*\,e^{i\kap}$. The expressions similar to Eqs.~\eqref{S1}-\eqref{S4} are also valid for the scalar products of the form $\skvv{\bar{v}}{v}$, $\skvv{v}{\bar{v}}$ and $\skvv{\bar{v}}{\bar{v}}$. To calculate these scalar products $x$ should be set to $\left(e^{i\bar{\kap}}\right)^* e^{i\kap}$, $\left(e^{i\kap}\right)^* e^{i\bar{\kap}}$ and $\left(e^{i\bar{\kap}}\right)^* e^{i\bar{\kap}}$, respectively. Expressions similar to Eqs.~\eqref{S5}-\eqref{S8} are valid for the calculation of $\skvk{\bar{v}}{v}$, $\skvk{v}{\bar{v}}$ and $\skvk{\bar{v}}{\bar{v}}$, if proper definition of $x$ is used. In the latter two cases $\ds{\kap}$ in Eqs.~\eqref{S5}-\eqref{S8} should be replaced by $\ds{\bar{\kap}}$.

Thus, in order to evaluate the Berry connection $A(k)$, we need to calculate the derivatives $\ds{C_j}$, $\ds{\bar{C}_j}$ ($j=\overline{1,4}$), $\ds{\eps}$, $\ds{\kap}$, $\ds{\bar{\kap}}$, 11 derivatives in total. Differentiating Eqs.~\eqref{C1}-\eqref{CD2} with respect to wave number $k$, we obtain the system of 10 linear equations containing 11 unknown derivatives. An additional (11-th) condition stems from the normalization of the periodic part of the doublon wave function:
\begin{align}
& \skvv{u_k}{u_k}=1\:,\label{N1}\\
& \frac{\partial}{\partial k}\,\skvv{u_k}{u_k}=0\:.\label{N2}
\end{align}
We also fix the gauge of the doublon wave function by the additional requirement
\begin{equation}\label{Gauge}
\text{Im}\,C_1=0\:.
\end{equation}
Then the calculation of Zak phase for doublon can be performed as follows:
\begin{itemize}
\item Find the coefficients $C_j$ and $\bar{C}_j$ following the procedure outlined in Sec.~I, using the normalization condition Eq.~\eqref{N1} and the gauge choice Eq.~\eqref{Gauge}.
\item Solve the system of 10 linear equations obtained by differentiation of Eqs.~\eqref{C1}-\eqref{CD2} expressing all the derivatives via $\ds{C_1}$. The latter derivative is purely real according to Eq.~\eqref{Gauge}.
\item Find $\ds{C_1}$ using the condition Eq.~\eqref{N2} and calculate all the remaining derivatives.
\item Calculate the scalar products Eq.~\eqref{S1}-\eqref{S8} and evaluate the Berry connection by Eqs.~\eqref{A1}, \eqref{A2}.
\item Find the Zak phase by means of the numerical integration of  the Berry connection over the entire Brillouin zone:
\end{itemize}

\begin{figure}[b]
\begin{center}
\includegraphics[width=0.9\linewidth]{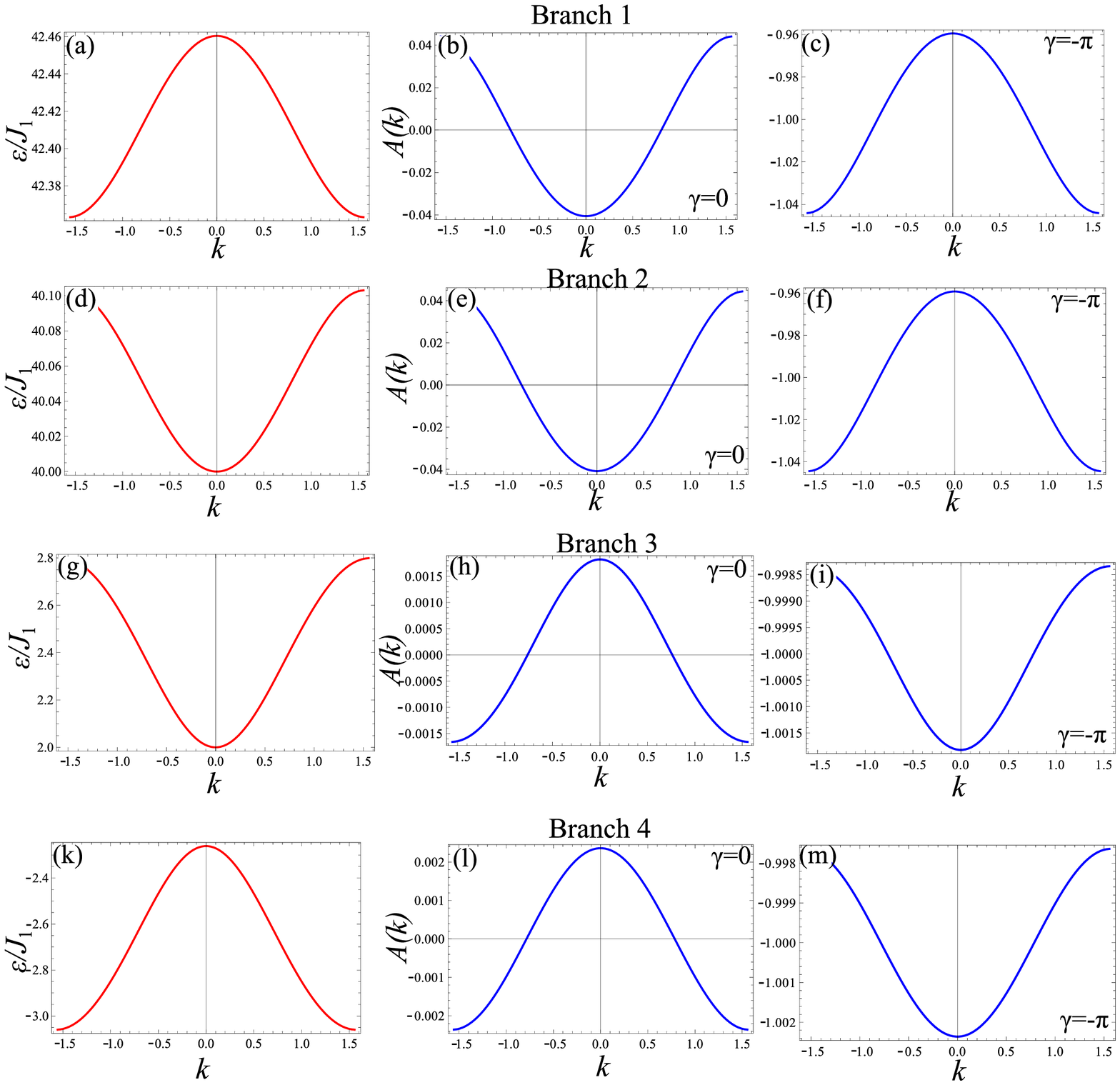}
\caption{Calculated doublon (a,d,g,k) dispersion and (b,c,e,f,h,i,l,m) Berry connection for  the dimer lattice with the ratio of tunneling constants equal to 5. (b,e,h,l) correspond to $J_1<J_2$ and (c,f,i,m) correspond to $J_1>J_2$. (a-c) First doublon zone, $U=20$. (d-f) Second doublon zone, $U=20$. (g-i) Third doublon zone, $U=1$. (k-m) Fourth doublon zone, $U=20$. Doublon zones are enumerated in the descending order with respect to their energy.}
\label{ris:Phase}
\end{center}
\end{figure}

\begin{equation}\label{ZakPhase}
\gamma=\int\limits_{-\pi/2}^{\pi/2}\,A(k)\,dk\:.
\end{equation}
The results of the calculation of the doublon dispersion and the Zak phase for all  four doublon bands are presented in Fig.~\ref{ris:Phase}. The upper and the lower doublon energy zones are characterized by the negative doublon effective mass, while the remaining two zones correspond to the positive doublon effective mass. Calculated Berry connection exhibits smooth dependence on the wave number for all doublon zones (besides the special case of doublon collapse, when the Zak phase loses its meaning). This allows us to calculate Zak phase by numerical integration of the Berry connection. According to our results, the Zak phase $\gamma=0$ if $J_1<J_2$ and $\gamma=\pi$ modulo $2\,\pi$, if $J_1>J_2$. Quite interestingly, this result is independent of the interaction strength $U$ as well as of the chosen doublon zone.

\end{document}